# Statistiques et visibilité des bibliothèques numériques : quelles stratégies de diffusion ?

Par Mathieu Andro(1), Gaëtan Tröger(2)

(1) INRA, DV IST, F-78026 Versailles, France. mathieu.andro@versailles.inra.fr
(2) Ecole des Ponts ParisTech, F-77455 Champs-sur-Marne, France.

## Résumé

Cet article compare les statistiques de consultation des principales bibliothèques numériques, interroge sur l'existence d'une relation entre le volume des entrepôts numériques et la visibilité sur Internet de chaque document numérisé, et analyse les conséquences des stratégies de diffusion des bibliothèques numériques françaises.
Les données statistiques ont été obtenues par des moyens classiques (enquête, littérature grise) et plus originaux (alexa.com, Google Trends)

En 2012, nous avons voulu mettre à l'épreuve l'hypothèse selon laquelle il existerait une forte corrélation entre la taille d'une bibliothèque numérique considérée sous l'angle du nombre de documents et la visibilité des documents sur Internet de chaque document numérisé. Une enquête sur les fonctionnalités et les statistiques des bibliothèques numériques a été menée, à partir de juillet 2012, sous la forme d'un questionnaire en ligne proposé au travers des listes de diffusion Bibliopat et ADBS aux professionnels des bibliothèques et de la documentation. Le faible taux de réponse n'a pas permis de tirer des conclusions solides de cette enquête. A ces premières données collectées, des statistiques provenant de comptes-rendus d'activités, de modules statistiques, et de sites spécialisés dans la webométrie ont également été utilisés.

## Statistiques de fréquentation issues du site alexa.com

Dans l'impossibilité d'obtenir systématiquement des données de consultation comparables des principales bibliothèques numériques, les données du site alexa.com ont été utilisées.

Alexa.com[1] est un service qui fait autorité dans la communauté des développeurs web. Ce site permet d'obtenir des estimations de la part relative du traffic web global des noms de domaines généraux qu'il est possible de comparer entre eux.
Une comparaison effectuée entre Gallica, Internet Archive, E-corpus, Europeana et Hathitrust[2] permet de constater une écrasante domination d'Internet Archive sur les autres grandes bibliothèques numériques, à l'exception probable de Google Books qui n'a pas pu être évalué car seuls les domaines de Google général auraient pu l'être[3]. Mais le biais présenté à propos de Google Books, peut sans doute également être invoqué à propos d'Internet Archive puisque cette base comprend par ailleurs la "wayback machine", à savoir une sauvegarde de différentes version de sites internets, un entrepôt de films, de fichiers musicaux ou audio qui ont sans contribué pour une très large part à l'afflux du public sur le site d'Internet Archive.

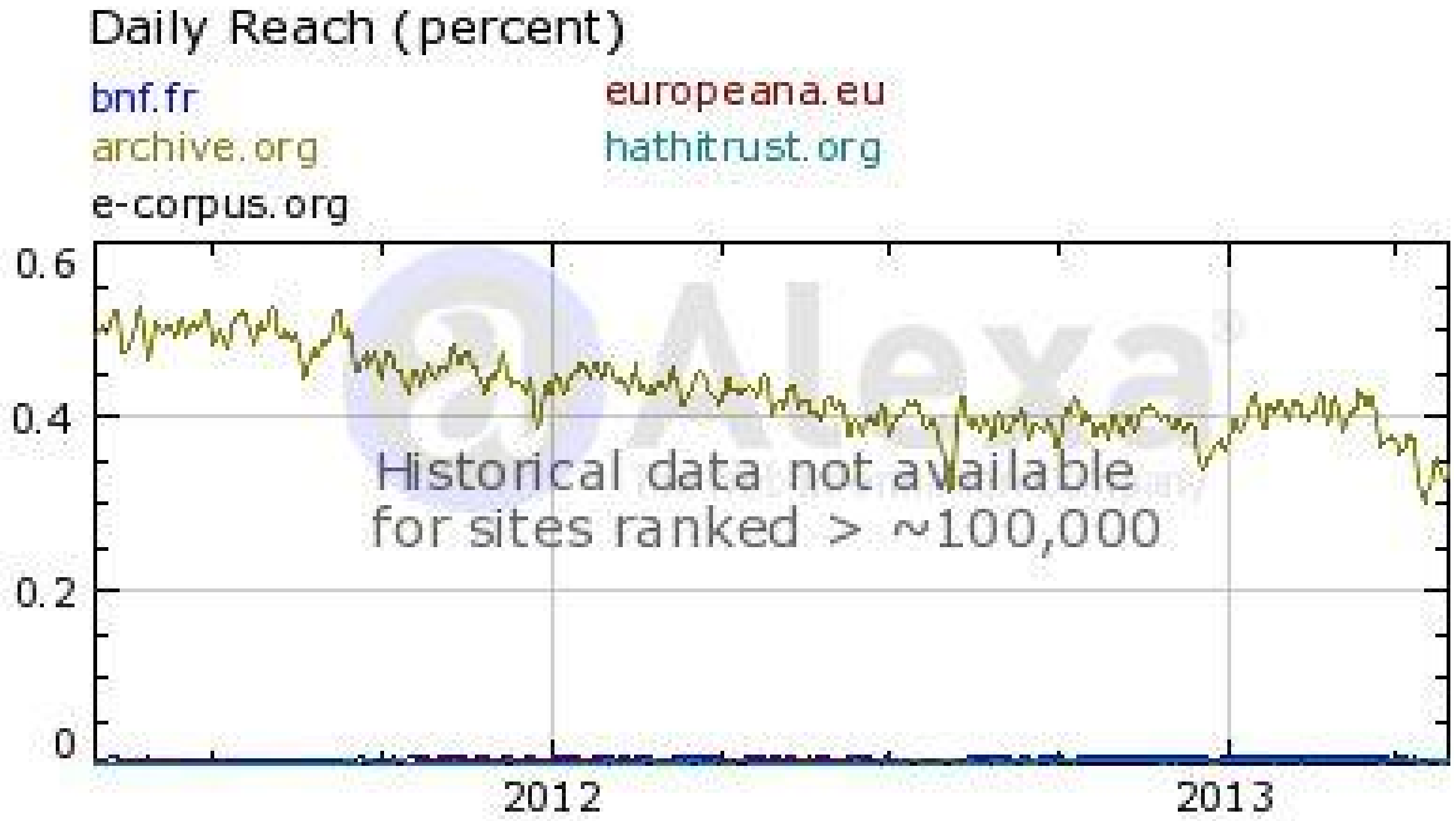

**Nombre relatif et comparé de visiteurs sur les noms de domaines bnf.fr, europeana.eu, archive.org, hathitrust.org et e-corpus.org d'après alexa.com depuis 2011[4]**

A présent, si on se concentre sur les autres bibliothèques numériques en dehors de Internet Archive, on observe une domination de Gallica (même si les données disponibles concernent l'ensemble du domaine bnf.fr) sur Europeana, le Hathi Trust et e-corpus :

[1] Alexa.com appartient à la société Amazon
[2] Les recherches ont respectivement portées sur les noms de domaine bnf.fr, archive.org, e-corpus.org, europeana.eu et hathitrust.org (date de consultation : .. / .. / 2013)
[3] De la même manière, les données portent sur bnf.fr et non sur la seule partie gallica.bnf.fr sur tout archive.org et non sur la seule partie consacrée aux livres numérisés.
[4] Bien que le nombre de visites sur archive.org soit passé de 227 244 392 en 2011 à 317 860 770 en 2012, la part de ce site sur le traffic web mondial a tendance à faiblement décliner.

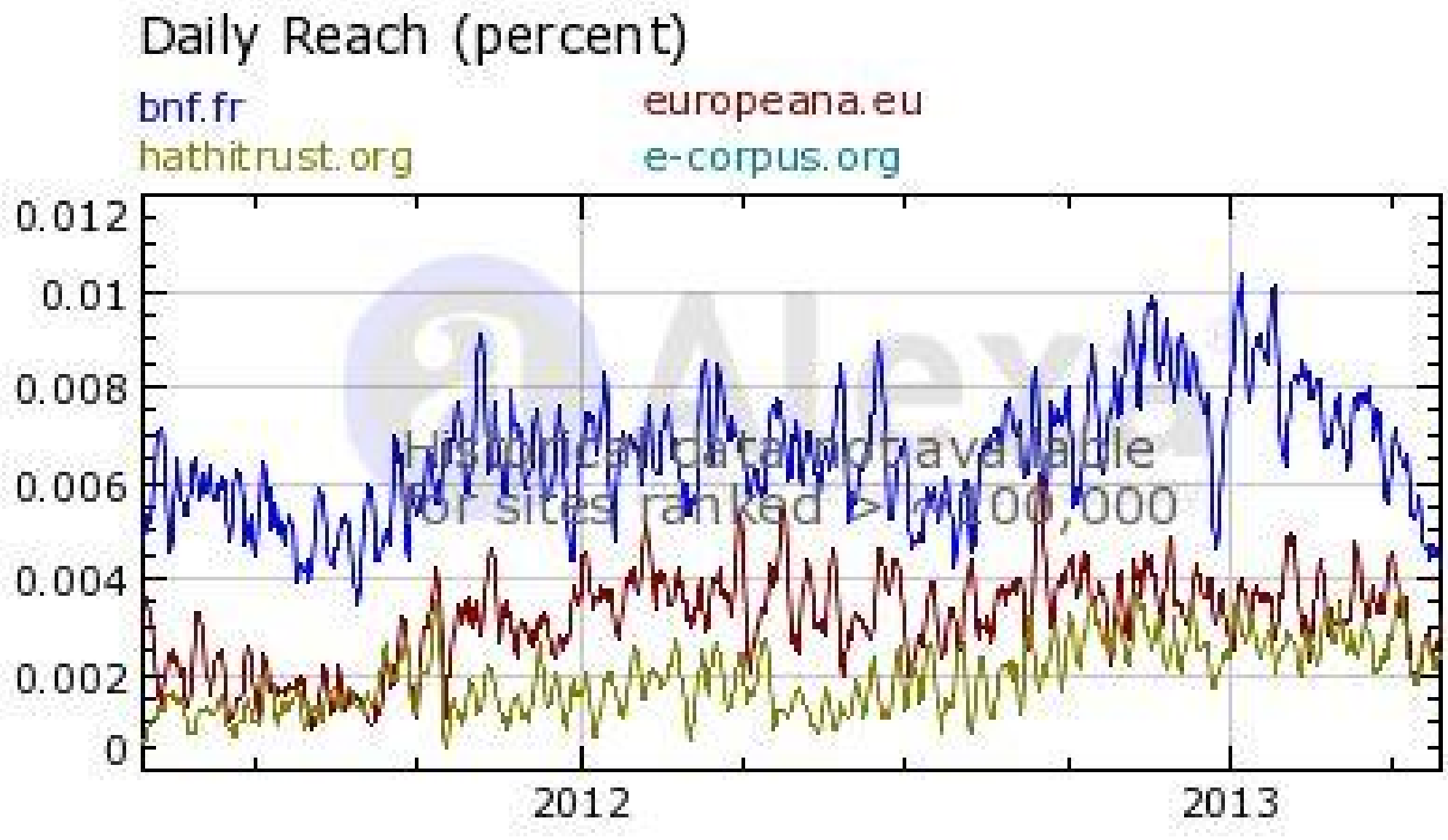

**Nombre relatif et comparé de visiteurs sur les noms de domaines bnf.fr, europeana.eu, hathitrust.org et e-corpus.org d'après alexa.com depuis 2011**

## Statistiques de saisie du nom des bibliothèques numériques dans Google

Le site Google Trends nous a permis d'obtenir les statistiques depuis 2004 du nombre relatif, sur une base 100, d'internautes ayant saisi tel ou tel mot clés dans le moteur de recherche Google. Concernant les bibliothèques numériques, Google Trends peut ainsi nous indiquer le nombre d'internautes qui ont saisi le nom de telle ou telle plateforme afin de s'y rendre. On observe ainsi que la domination de Internet Archive depuis 2004 est dépassée par Google Books à partir de 2007, mais que Internet Archive conserve sa 2ème place, devant Gallica et Europeana, eux mêmes devant le Hathi Trust qui peine à trouver son public.

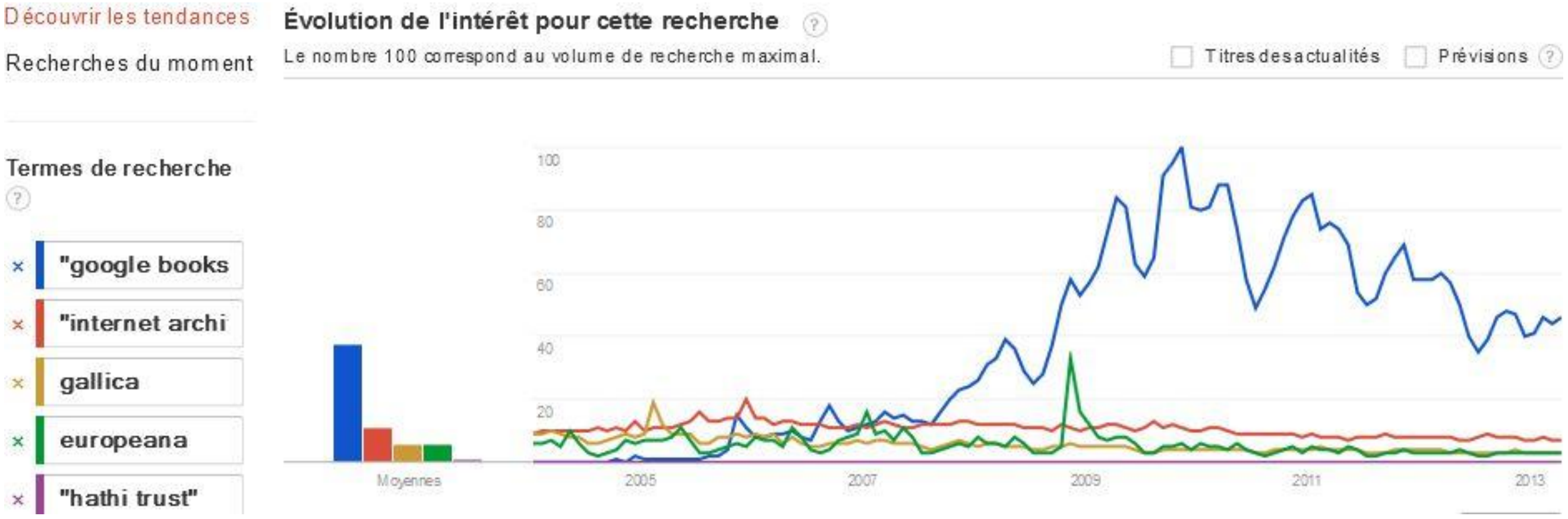

**Nombre relatif et comparé de saisies depuis 2004, dans le monde et dans le moteur de recherche Google des expressions "google books", "internet archive"+"archive.org", gallica, europeana et "hathi trust" d'après Google Trends**

Néanmoins, si on étudie exclusivement les internautes de France et non plus ceux du monde entier, les résultats sont sensiblement différents. Gallica arrive, cette fois, en première position, devant Google Books qui progresse depuis 2007, puis Internet Archive qui semble connaître de récents sursauts et Europeana. La Bibliothèque nationale de France semble donc conserver la première place pour ce qui concerne la satisfaction des besoins en livres numérisés des internautes de France.

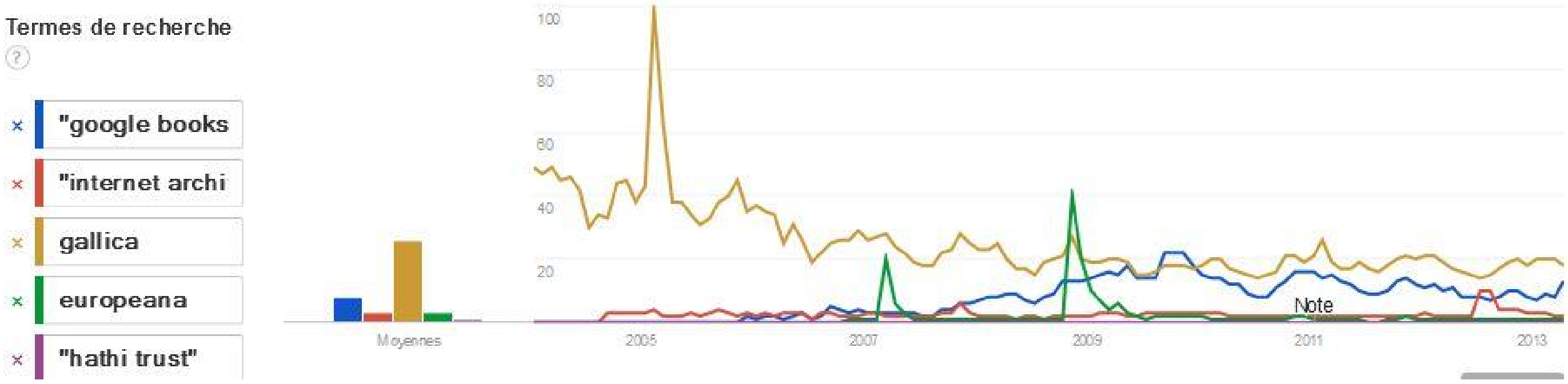

**Nombre relatif et comparé de saisies depuis 2004, en France et dans le moteur de recherche Google des expressions “google books”, “internet archive”+”archive.org”, gallica, europeana et “hathi trust” d’après Google Trends**

Le pic qu’on peut observer pour Gallica en février 2005 pourrait faire suite, à la campagne médiatique menée par Jean-Noël Jeanneney, alors Directeur de la Bibliothèque nationale de France, en réaction au projet Google Print de numériser 15 millions de livres. Ce projet, qui prendra le nom de Google Books, numérisera d’ailleurs un nombre de livres bien supérieur aux objectifs annoncés puisque plus de 20 millions de livres ont déjà été numérisés.

Le pic de mars 2007 qui concerne Europeana reflète probablement la mise en ligne de la pré version d’Europeana à l’occasion du salon du livre qui s’était tenu à Paris le 22 mars 2007. Celui de novembre 2008 correspondrait, quant à lui, au lancement médiatisé mais difficile de Europeana.

# Relation entre le volume des grandes bibliothèques numérique et leur visibilité sur le web

Il est impossible de comparer le traffic web d’une plateforme diffusant 20 millions de documents avec celui d’une petite bibliothèque numérique qui n’en diffuse que 500. Afin de pouvoir comparer des statistiques de consultation entre des bibliothèques numériques de tailles très différentes, nous avons donc rapporté le nombre de leurs visiteurs au nombre de documents qu’elles ont mis en ligne et décidé de considérer le nombre moyen de visiteurs uniques par document numérique et par mois. Les statistiques de 2012 n’étant pas encore toujours disponibles, seules les données 2011 ont été comparées.

| | Nombre de livres | Nombre de visites ou de téléchargements en 2011 | Nombre de visites ou de téléchargements par mois en 2011 | Nombre de visites ou de téléchargements par mois et par livre |
|---|---|---|---|---|
| Internet Archive[5] | 3 678 804 livres (au 25 octobre 2012) | 227 244 392 téléchargements | 18 937 033 téléchargements par mois | 5,15 téléchargements par livre et par mois minimum[6]. |
| Gallica[7] | 224 322 livres parmi 1,6 millions de documents (fin 2011) | 9 485 603 visites | 790 467 visites par mois | 3,52 visites par livre et par mois 0,49 visites par document et par mois |
| e-corpus | 23 870 textes (au 1er janvier 2012) | 418 215 vistes | 34 851 visites par mois | 1,46 visites par livre et par mois minimum |
| Bibliothèques numériques autonomes[8] | 308 livres<br><br>1193<br>4030<br>8592<br>1424<br>1197<br>264 | 7200 visites<br><br>23628<br>26036<br>77390<br>33782<br>5751<br>16771 | 600 visites par mois<br>1969<br>2170<br>6449<br>2815<br>479<br>1398 | 1,95 visites par livre et par mois<br>1,65<br>0,54<br>0,75<br>1,98<br>0,4<br>5,29 |

A partir d'un certain seuil, il semble que plus une bibliothèque numérique diffuse de livres numérisés, plus le nombre moyen de visites par mois et par livre soit important.

# Les stratégies de diffusion des bibliothèques numériques de France

Il est fort probable qu'en groupant les collections de plusieurs institutions, et en présentant sur le web une masse plus importante de documents, les bibliothèques numériques collectives et

[5] d'après http://indexer.us.archive.org/services/collection-downloads.php?mediatype=texts

[6] Nous avons divisé le nombre de downloads de 2011 par le nombre de livres mis en ligne au 25 octobre 2012, un nombre supérieur à celui du 31 décembre 2011 que nous aurions utilisé si nous l'avions trouvé et qui aurait donné un nombre de downloads par mois et par document supérieur à 5,15.

[7] d'après http://webapp.bnf.fr/rapport/html/numerique/1_coope_num.htm

[8] D'après les résultats de l'enquête menée en 2012. Les bibliothèques concernées ne sont pas citées à la demande de certaines d'entre elles. Quelques données statistiques ont été obtenues par extrapolation car les informations fournies portaient sur des périodes incomplètes.

mutualisées soient plus attractives pour les usagers. Leur PageRank, leur visibilité et leur référencement pourraient également être meilleurs pour les moteurs de recherche car ceux-ci tiennent compte du nombre de liens qui pointent vers tel ou tel nom de domaine et ce nombre de liens est potentiellement bien plus important dans le cas d'une grosse bibliothèque numérique collective que dans celui d'un petit site autonome. Il y aurait ainsi une corrélation entre les statistiques de consultation et le nombre de documents mis en ligne, une corrélation qui favoriserait les stratégies de mutualisation à une échelle internationale.

Le moissonnage des métadonnées des bibliothèques numériques en OAI-PMH permet déjà, aux petites bibliothèques numériques, une exposition de leurs propres documents numérisés sur la "bibliothèque numérique nationale" Gallica et une visibilité accrue par l'afflux des visites en provenance de Gallica. Mais cela n'est possible qu'à condition qu'elles disposent d'un moyen de diffusion et d'une exposition OAI de leurs notices. Or, c'est souvent l'objet de difficultés pour les bibliothèques. En effet, les institutions renoncent encore trop souvent à diffuser ce qu'elles ont numérisé car le développement d'une bibliothèque numérique autonome nécessite des moyens humains et financiers importants pour des résultats en fonctionnalités et en visibilité souvent décevants. Quasiment inexistantes sont, par exemple, les bibliothèques qui autorisent la lecture de leurs livres numérisés sur liseuses aux format ePub ou mobi pour Kindle comme cela est proposé, par exemple, systématiquement, sur Internet Archive[9]. Or il est presque impossible de lire un livre in extenso sur un écran d'ordinateur. Ainsi, nombreux sont les projets qui, négligeant les besoins des usagers et les pratiques des internautes, ne présentent qu'une offre de vitrine superficielle, limitée et ne répondant bien souvent qu'à une logique de communication politique ou institutionnelle.

Du strict point de vue de la conservation, si une majeure partie des bibliothèques françaises renonce encore à diffuser sur Internet les documents qu'elles ont pourtant numérisés avec des budgets conséquents, cela signifie aussi que nombreux sont les documents numérisés qui demeurent ainsi sur des supports dont la durée de conservation est très limitée et qui risquent, de surcroît, d'être numérisés en doublon. Ceux qui sont diffusés en ligne le sont souvent sur des plateformes locales dont la maintenance est éphémère et dont la pérennité est très fragile. Mais, Gallica ne peut offrir, pour le moment, un débouché à ces numérisations, puisqu'il n'est, malheureusement, toujours pas possible d'y "*déposer*" ses fichiers numérisés comme il est possible de le faire sur Internet Archive ou sur e-corpus.

# Conclusion

Si la première des bibliothèques de France respecte le droit des autres bibliothèques à disposer d'elles-mêmes, elle est néanmoins, consciente des difficultés dans lesquelles se trouvent pour la plupart d'entre elles. C'est la raison pour laquelle, un projet "Gallica marque blanche" est en cours de développement et devrait offrir cette possibilité à la rentrée 2013 avec, pour commencer, la Bibliothèque Nationale et Universitaire de Strasbourg.
Disposer de Gallica en marque blanche devrait apporter des fonctionnalités et une visibilité optimales aux bibliothèques et leur offrir probablement l'une des meilleures solutions de diffusion de leurs numérisations. Ce projet aura nécessité que la BnF accepte d'ouvrir, aux bibliothèques extérieures, son catalogue sur lequel Gallica est adossé. Il est donc probablement le fruit d'efforts soutenus de la part des cadres dirigeants de la BnF afin de conduire un changement non négligeable dans la culture de l'institution pour qu'elle accepte d'intégrer des métadonnées produites par des catalogueurs extérieurs.

[9] Il s'agit de fichiers réalisés à partir d'un OCR souvent brut dont la qualité reste donc variable. Néanmoins, les documents du projet Gutenberg, qui participe à Internet Archive, proposent, par exemple, un OCR d'excellente qualité, en grande partie grâce à la correction participative (*crowdsourcing*) du site Distributed Proofreaders.

# Remerciements

Nous remercions Catherine Thiolon des éditions Quae à l'Institut National de la Recherche Agronomique (INRA) pour sa relecture et ses remarques.

# Bibliographie